\documentclass[12pt]{article}
\usepackage{amsmath,amssymb}
\usepackage{graphicx,color}
\usepackage[breaklinks]{hyperref}
\usepackage{cite,epsf}
\usepackage[bf]{caption}

\newcommand{\rr}{\mathbb{R}}

\newcommand{\sz}{\mathbb{S}}

\newcommand{\be}{\begin{equation}}
\newcommand{\ee}{\end{equation}}
\newcommand{\ba}{\begin{aligned}}
\newcommand{\ea}{\end{aligned}}
\newcommand{\ben}{\begin{displaymath}}
\newcommand{\een}{\end{displaymath}}
\newcommand{\bea}{\begin{eqnarray}}
\newcommand{\eea}{\end{eqnarray}}
\newcommand{\bean}{\begin{eqnarray*}}
\newcommand{\eean}{\end{eqnarray*}}
\newcommand{\f}{\frac}
\newcommand{\p}{\partial}

\usepackage{amsmath,amssymb}
\usepackage{graphicx,color}

\def\th {\theta}
\def\a {\alpha}

\def\b {\beta}

\def\g {\gamma}
\def\G {\Gamma}
\def\d {\delta}
\def\e {\epsilon}
\def\s {\sigma}
\def\e {\epsilon}
\def\z {\zeta}

\def\m{\mu}
\def\n{\nu}

\definecolor{green}{rgb}{0,0.5,0}

\def\p{\partial}

\usepackage{epsfig}

\begin{document}

\begin{titlepage}       \vspace{10pt} \hfill 

\vspace{20mm}

\begin{center}

{\large \bf Recursive relations for the S-matrix of Liouville theory}

\vspace{30pt}

George Jorjadze$^{a,\,b}~$, Lado Razmadze$^{c}$ and Stefan Theisen$^{d}$ 
\\[6mm]

{\small
{\it ${}^a$Free University of Tbilisi,\\
	240 Davit Agmashenebeli Alley, 0159, Tbilisi, Georgia}\\[2mm]
{\it ${}^b$Razmadze Mathematical Institute of TSU,\\
2 Merab Aleksidze II Lane,  Tbilisi 0193, Georgia}\\[2mm]
{\it ${}^c$Forschungszentrum J\"ulich GmbH,\\
Wilhelm-Johnen-Stra\ss e, 52428, J\"ulich, Germany}\\[2mm]
{\it ${}^d$Max-Planck-Institut f\"ur Gravitationsphysik, Albert-Einstein-Institut,\\ 
14476, Golm, Germany}
}

\vspace{20pt}

\end{center}

\centerline{{\bf{Abstract}}}
\vspace*{5mm}
\noindent

We analyze the relation between the vertex operators of the in and out fields in Liouville
theory. This is used to derive equations for the S-matrix, from which a recursive
relation for the normal symbol of the S-matrix for discrete center-of-mass momenta is obtained. 
Its solution is expressed as multiple contour-integrals of a generalized
Dotsenko-Fateev type. This agrees with the functional integral representation of the
scattering matrix of Liouville theory which we had proposed previously.

\vspace{15pt}

\end{titlepage}


\section{Introduction and summary}

Liouville theory has been for many years the prime example of an interacting two-dimen\-sional 
field theory where 
exact results can be obtained. It is an integrable conformal field theory with the remarkable   
property that conformal invariance and integrability can be maintained in the quantum theory. 
For thorough reviews we refer to \cite{Nakayama,Teschner}  and for more recent developments 
to \cite{Kupi1,Kupi2,Witten}.  

In this note we consider quantum Liouville theory on a Minkowski cylinder\footnote{In contrast to the 
strip, there are no bound states on the cylinder.}, more specifically its 
scattering matrix. The solutions of the theory are such that at asymptotic times they approach 
massless free fields and the scattering matrix is the overlap between asymptotic $in$- and $out$-states. 
There are various ways of computing exact $S$-matrix elements. They use the integrability of the system, 
which implies the existence of an infinite number of conserved charges which can be combined 
into the Virasoro-modes of the energy-momentum tensor.  At asymptotic times 
it takes the form of the (improved) energy-momentum tensor of a free massless scalar field. 

We have studied the $S$-matrix previously in refs.\cite{JT2,JT3}. In particular we have proposed a functional integral 
representation for the generating functional of $S$-matrix elements 
which involves a non-local, but nevertheless simple one-dimensional classical action of a scalar field 
on a circle. 
Note that a similar functional integral representation of the $S$-matrix can be realized for the SL(2,$\rr$)/U(1) coset WZW model 
and SL(3,$\rr$) Toda theory \cite{DJM}.
In Liouville theory we have checked this functional integral representation in various ways. 
First in perturbation theory at one- and two-loop order and then also to all orders in $\hbar$, but the latter only for discrete (imaginary) values 
of the center-of-mass momentum.  
In this case the functional integral collapses to a finite dimensional integral, the dimension depending on the 
discrete momentum value.  In this note we will rederive these finite dimensional integrals 
in an independent way which uses only the known relation between the asymptotic $in$- and 
$out$-fields.  The explicit evaluation of these integrals beyond what was presented in \cite{JT3} 
remains, however, an open problem. 

\section{Asymptotic fields of Liouville theory}

In the first part of this section we introduce the asymptotic fields of the classical Liouville theory 
and analyze the relation between the $in$- and $out$-fields, briefly summarizing some results of \cite{JT2, JT3}. 
In the second part of the section we introduce exponentials of the asymptotic field operators as primary fields and
present the quantum analog of the classical relation between the asymptotic fields. 

\subsection{Classical asymptotic fields}

We study the scattering problem in Liouville theory on a cylindrical spacetime with coordinates $\s\in \sz^1$ and $\tau\in \rr^1$. 
The dynamics of the system is governed by Liouville's equation
\be\label{L-eq}
\p_{\bar x}\p_x\Phi+\m^2\,{e}^{2\Phi}=0\,,
\ee
where $x=\tau+\s$ and $\bar x=\tau-\s$ are the chiral coordinates and $\m^2$ is a positive constant.

At asymptotic times, $\tau\rightarrow \mp\infty$,  the Liouville field becomes a free massless field. 
The asymptotic fields have the standard Fourier mode expansions 
\be\ba\label{in-out fields}
&\Phi_{\text{in}~}=q+p\,\tau+{i}\sum_{n\neq 0}\left(\f{a_n}{n}\,{e}^{-{i}nx} +\f{\bar a_n}{n}\,e^{-{i}n\bar x}\right)\,,
\\[1mm]
&\Phi_{\text{out}}=\tilde q+\tilde p\,\tau+{i}\sum_{n\neq 0}\left(\f{b_n}{n}\,{e}^{-{i}nx} +\f{\bar b_n}{n}\,{e}^{-{i}n\bar x}\right)\,,
\ea\ee
which can be written as a sum over chiral and anti-chiral parts. 
It is important to note that the $out$-field always differs from the $in$-field. In particular, the momentum 
zero mode of the $in$-field is positive, $p>0$, while $\tilde p=-p<0$, similarly to the particle dynamics 
in the exponential potential, which corresponds to the homogeneous Liouville field $\p_\s\Phi(\tau,\s)=0$. 

By the general solution of the Liouville equation one can find the explicit form of the exponential of the 
$out$-field $e^{-\Phi_{\text{out}}}$ in terms of the $in$-field
\be\label{out=in}
e^{-\Phi_{\text{out}}(x,\bar x)}=\m^2\,e^{-\Phi_{\text{in}}(x,\bar x)}\,\int_{-\infty}^{x}dy\int_{-\infty}^{\bar x}d\bar{y}\,e^{2\Phi_{\text{in}}(y,\bar{y})}\,.
\ee

Another remarkable relation is the form of the Liouville field exponential $e^{-\Phi}$  in terms of the exponentials of the asymptotic fields 
\be\label{e^-Phi}
e^{-\Phi(x,\bar x)}=e^{-\Phi_{\rm in}(x,\bar x)}+e^{-\Phi_{\rm out}(x,\bar x)}\,.
\ee
This relation, together with \eqref{out=in}, defines a parameterization of the interacting field $\Phi$ 
in terms of the asymptotic free field $\Phi_{\rm in}$.

If we switch off the oscillator modes and consider the homogeneous $in$-field consisting of only the zero-modes
$\Phi_{\rm in}=q+p\tau$, 
then by \eqref{out=in}, the $out$-field is also homogeneous, 
$\Phi_{\rm out}=\tilde q+\tilde p\tau$, with $\tilde q=-q+\log(p^2/\m^2)$ and $\tilde p=-p$. 
Therefore, if all $a$-modes vanish, then the $b$-modes are also in the vacuum sector.

One can similarly parameterize $\Phi$ by the $out$-field, using the expression of the $in$-field in terms of the $out$-field, obtained by the inversion of \eqref{out=in}
\be\label{in=out}
e^{-\Phi_{\text{in}}(x,\bar x)}=\m^2\,e^{-\Phi_{\text{out}}(x,\bar x)}\,\int_x^{\infty}dy\int_{\bar x}^{\infty}d\bar{y}\,e^{2\Phi_{\text{out}}(y,\bar{y})}\,.
\ee 
Note that the conditions $p>0$ and $\tilde{p}<0$ provide convergence of the integrals in \eqref{out=in} and \eqref{in=out}, respectively. Below we use \eqref{out=in}.

In the Hamiltonian formulation Liouville theory has the standard canonical structure and  the parametrization of the 
Liouville field in terms of the $in$ (or the $out$) field induces the canonical structure for the asymptotic fields.
Specifically, the analysis of the canonical
symplectic form of Liouville theory shows that the Fourier modes of the asymptotic fields
satisfy the canonical Poisson brackets relations\cite{JT2} 
\bea\label{in-canonical PB}
&&\{p,q\}=1\,, 
\quad \{a_m,a_n\}=\f{{i}}{2}\,m\d_{m,-n}\,,\quad \{\bar a_m,\bar a_n\}=\f{{i}}{2}\,m\d_{m,-n}\,,\\[1mm] \label{out-canonical PB}
&&\{\tilde p,\tilde q\}=1\,, \quad
 \{b_m,b_n\}=\f{{i}}{2}\,m\d_{m,-n}\,,\quad ~\{\bar b_m,\bar b_n\}=\f{{i}}{2}\,m\d_{m,-n}\,.
\eea
One can also check by direct computation that (7) follows form (3) and (6).

Indeed, the canonical Poisson brackets \eqref{in-canonical PB} are equivalent to the causal relation between the $in$-field exponentials
\be\label{PB E's}
\{e^{-\Phi_{\text{in}}(x,\bar x)},e^{-\Phi_{\text{in}}(y,\bar y)}\}=\f \pi 2\,\left[\e(x-y)+\e(\bar{x}-\bar{y})\right]\,e^{-\Phi_{\text{in}}(x,\bar x)}\,e^{-\Phi_{\text{in}}(y,\bar y)},
\ee
where
\be\label{eps(x)=}
\e(x)=\f 1 \pi\left(x+i\sum_{n\neq 0}\f{e^{-inx}}{n}\right),
\ee
is the stair-step function: $\e(x)=2N+1,~$ for $~2\pi N< x < 2\pi(N+1)$. In particular 
at equal $\tau$ this Poisson bracket vanishes. 
In Appendix A we show how from \eqref{out=in} and \eqref{PB E's} one derives that  
the $out$-field exponentials satisfy the the same causal Poisson brackets.
This proves that the map \eqref{out=in} is canonical.

Taking into account 
 \eqref{A=sinh} and its anti-chiral counterpart, we represent the map \eqref{out=in} as
\be\label{out=in1}
e^{-\Phi_{\text{out}}(x,\bar x)}=\m^2\,e^{-\Phi_{\text{in}}(x,\bar x)}\,\f{1}{4\sinh^2(\pi p)}\int_0^{2\pi}dy\int_0^{2\pi}d\bar{y}\,e^{2\Phi_{\text{in}}(y+x,\bar{y}+\bar{x})-2\pi p}\,.
\ee
We will use this relation in the quantum theory to fix the operator ordering ambiguities.

With help of eq. \eqref{out=in} we can express the Fourier modes of the $out$-field in terms of those of the $in$-field
\be\ba\label{q=}
&\tilde p=-p\,,& \qquad &\tilde q=-q-\log{\m^2}- Q_0(p,a, a^*)-\bar Q_0(p,\bar a,\bar a^*)&\\[1mm]
&b_n=a_n+in Q_n(p,a,a^*)\,,&  &\bar b_n=\bar a_n+in \bar Q_n(p,\bar a,\bar a^*)\,,&\, 
\ea\ee
where
\bea\label{Q=}
&&Q_n(p,a,a^*)= \int_0^{2\pi}\f{dx}{2\pi}\,e^{inx}\log A_p(x)\,, \quad 
\bar Q_n(p, \bar a,\bar a^*)=\int_0^{2\pi}\f{d\bar x}{2\pi}\,e^{in\bar x}
\log \bar A_p(\bar x)\,, \\[1mm]
\label{A_p(x)}
&& A_p(x)=\int_{-\infty}^{0}\,dy \,e^{py+2a(y+x)+2a^*(y+x)}, \qquad 
\bar A_p(\bar x)=\int_{-\infty}^{0}d\bar y\,e^{p\bar y+2\bar a(\bar y+\bar x)+2\bar a^*(\bar y+\bar x)},
\eea
$a(x)$, $\bar{a}(\bar{x})$ are the positive frequency parts of the chiral and anti-chiral fields
\be\label{a(x)}
a(x)={i}\sum_{k> 0}\f{a_k}{k}\,{e}^{-{i}kx}, \qquad \bar{a}(\bar{x})={i}\sum_{m> 0}\f{\bar a_k}{k}\,e^{-{i}k\bar x},
\ee
while the complex conjugates $a^*(x)$, $\bar{a}^*(\bar{x})$ are their negative frequency parts.

Thus, the ${b}$-modes depend on $a$-modes but not on the  $\bar a$-modes.  Similarly,  
$\bar b$-modes depend on $\bar a$'s but not on $a$'s. This is an indication 
for the chiral structure of the $S$-matrix.

Finally note that 
\be\label{b=a}
Q_n(p,a,a^*) \rightarrow 0\,, \qquad \text{when} \quad p\rightarrow 0
\ee
and similarly for $\bar{Q}_n$.  It means that scatterings disappears at $p=0$. 
This fact follows from \eqref{Q=} and \eqref{A_p(x)}, where it is convenient to use the representation of $A_p(x)$ similar to \eqref{A=sinh}.
We will use this fact in Section 4 to find recursive relations for the $S$-matrix.

\subsection{Quantum asymptotic fields}

We apply canonical quantization and use the standard `hat' notation for the operators 
corresponding to the classical counterparts.

Based on the classical picture, one has the canonical commutation relations
\be\ba\label{canonical CR}
&[\hat q,\hat p]=i\hbar~, 
\quad &[\hat a_m,\hat a_n]=\f{\hbar}{2}\,m\d_{m,-n}~,\quad &[\hat{\bar a}_m,\hat{\bar a}_n]=\f{\hbar}{2}\,m\d_{m,-n}~;\\[1mm]
&[\hat{\tilde q},\hat{\tilde p}]=i\hbar~, 
 &[\hat b_m,\hat b_n]=\f{\hbar}{2}\,m\d_{m,-n}~,\quad &[\hat{\bar b}_m,\hat{\bar b}_n]=\f{\hbar}{2}\,m\d_{m,-n}~,
\ea\ee
together with
\be\label{zero CR}
[\hat a_m,\hat{\bar b}_n]=[\hat{\bar a}_m, \hat b_n]=0~.
\ee

However, the commutation relations of the $a$-modes with the $b$-modes (and  $\bar a$'s with $\bar b$'s) 
are highly non-trivial and they are defined by the quantum version of  \eqref{out=in1},
which has the following form 
\be\label{out=in_q}
:e^{-\hat\Phi_{\text{out}}(x,\bar x)}:=\f{\m_q^2}{2\sinh(\pi\hat p)}\,:e^{-\hat\Phi_{\text{in}}(x,\bar x)}:\int_0^{2\pi}dy\int_0^{2\pi}d\bar{y}\,:e^{2\hat\Phi_{\text{in}}(y+x,\bar{y}+\bar{x})-2\pi\hat p}:
\,\f{1}{2\sinh(\pi\hat p)}\,,
\ee
with
\be\label{mu_q=}
\m_q^2=\m^2\,\f{\sin(\pi\hbar)}{\pi\hbar}~.
\ee
The normal ordered free-field exponentials in \eqref{out=in_q} are the vertex operators 
defined in a standard way by separating 
the  zero mode, the positive and the negative frequency parts, as e.g. in 
\be\ba\label{N-exponents}
&:e^{-\hat\Phi_{\text{out}}(x,\bar x)}:=e^{-\hat{\tilde q}+\hat p\tau}\,e^{-\hat{\bar b}^\dag(\bar x)}\,e^{-\hat b^\dag(x)}\,e^{-\hat{\bar b}(\bar x)}\,e^{-\hat b(x)}\,,\\[1mm]
\ea\ee
where the operators $\hat b(x)$, $\hat{\bar b}(\bar x)$ are given similarly to \eqref{a(x)}
\be\label{exp op}
\hat b(x)=i\sum_{k>0}\f{\hat b_k}{k}\,{e}^{-ikx},\qquad \hat{\bar b}(\bar x)=
i\sum_{k>0}\f{\hat{\bar b}_k}{k}\,{e}^{-ikx},
\ee
and $\hat b^\dag(x)$, $\hat{\bar b}^\dag(\bar x)$ are their Hermitian conjugates. 

The transition from the classical expression \eqref{out=in1} to the quantum expression \eqref{out=in_q}
is not unique due to operator ordering ambiguities. 
The requirements which lead from \eqref{out=in1} to \eqref{out=in_q} are:
\begin{itemize}

\item[a)] The spectrum of $\hat p$ is positive and $\hat{\tilde p}=-\hat p$, as in the classical theory.

\item[b)] The relation \eqref{e^-Phi} holds in quantum theory without deformation.
This assumption is in accordance with the previous one, since the asymptotic behavior of the $in$- and the $out$-field 
exponentials are defined by the factors $e^{-p\tau}$ and $e^{p\tau}$, respectively.

\item[c)] The conformal weight of $e^{2\,\hat \phi_{\text{in}}(z,\bar z)}$ is $(1,1)$. 
Then, the integration in \eqref{out=in_q} gives a conformal scalar and, 
therefore, the $in$- and $out$-field exponentials have the same conformal weights which, due to 
\eqref{e^-Phi}, coincides with the weight of $e^{-\hat\Phi}$. 

\item[d)] The exponentials of the asymptotic fields are Hermitian. 
This follows from 
\be\label{comm=0}
:e^{-\hat\Phi_{\text{in}}(x,\bar x)}:\,:e^{2\hat\Phi_{\text{in}}(x+y,\bar{x}+\bar{y})-2\pi p}:\,=
:e^{2\hat\Phi_{\text{in}}(x+y,\bar{x}+\bar{y})-2\pi p}:\,:e^{-\hat\Phi_{\text{in}}(x,\bar x)}:
\ee
which holds for $y\in(0,2\pi)$ and $\bar{y}\in(0,2\pi)$.

\item[e)]  Conformal symmetry and Hermiticity do not fix the $\hat p$-dependent factors, since $\hat p$ is a conformal scalar.
They are determined via the locality condition for the Liouville field
\be\label{locality}
[e^{-\hat\Phi(\tau,\s)}, e^{-\hat\Phi(\tau,\s')}]=0\,.
\ee
With the chosen ordering they are the same as in the classical expression \eqref{out=in1}.

\item[f)] 
Finally, the requirement that the quantum Liouville field satisfies equation \eqref{L-eq} determines 
the overall factor as in \eqref{mu_q=}.

\end{itemize}

For the technical details related to these requirements and how they lead from \eqref{out=in1} to \eqref{out=in_q},  
we refer to  refs.\cite{Braaten:1982yn, OW, Teschner, Dorn:2008sw}.

\section{Equation for the $S$-matrix}

In this section we introduce the coherent states of the asymptotic fields as the generators for the $in$ and $out$ 
Fock spaces and use the operator relation \eqref{out=in_q} to derive the equation for the transition amplitude 
from the coherent $in$ state to the coherent $out$ state. This defines the $S$-matrix  of the theory. 
We then show how to obtain the reflection amplitude and Fock space transition amplitudes.. 

Consider the momentum dependent asymptotic vacuum states: $|p\rangle_{\text{in}}$, with $p>0$, and
$|\tilde p\rangle_{\text{out}}$, with $\tilde p< 0$.
Motivated by the classical situation 
we \textit{assume} that the states $|p\rangle_{\text{in}}$ are annihilated  not only  by $\big(\hat a(x)$, $\hat{\bar a}(\bar x)\big)$, 
but also by $\big(\hat b(x)$, $\hat{\bar b}(\bar x)\big)$, and likewise for $|\tilde p\rangle_{\text{out}}$. 
Therefore, these states are related by
\be\label{Q p_in=-p_out}
|p\rangle_{\text{in}}=R(p)\, |-p\rangle_{\text{out}}\,, 
\ee
where $R(p)$ is called the reflection amplitude, and one has
\be\label{R(p)}
_{\text{out}}\langle \tilde p\,|\,p\rangle_{\text{in}}=\d(\tilde p+p)\,R(p)\,.
\ee
Since apart from the zero modes the two chiral sectors decouple, it suffices to discuss one of them; 
the other then follows immediately. 

We introduce the asymptotic coherent states
\be\label{CS}
|p, a\rangle_{\text{in}} =\exp\left(\f{2}{\hbar}\sum_{k>0}\f{a_k}{k}\,\hat a_k^\dag\right)|\,p\rangle_{\text{in}}~,
\qquad _{\text{out}}\langle \tilde p,\,b^*| =_{\text{out}}\langle \tilde p\,|\exp
\left(\f{2}{\hbar}\sum_{k>0}\f{b^*_k}{k}\,\hat b_k\right)~,
\ee
and the corresponding transition amplitude 
$_{\text{out}}\langle \tilde p,\,b^*|p, a\rangle_{\text{in}}$. 
Due to $\hat{\tilde p}+\hat p=0$ it has the structure 
\be\label{S-matrix}
_{\text{out}}\langle \tilde p,\,b^*|p, a\rangle_{\text{in}}=\,_{\text{in}}\!\langle \tilde p,b^*|\hat S|p,a\rangle_{\text{in}}
=\d(\tilde{p}+p)\,R(p)\,S_p[b^*,a]~,
\ee
where $\hat S$ is the scattering operator. 
$S_p[b^*, a]$ is  a holomorphic function of the modes $(b^*_k, a_k)$, for $k>0$ and contains $p$ as a parameter. 
Note that from \eqref{R(p)} it follows that  $S_p[0,a]=S_p[b^*,0]=1$. 

We will now derive equations for $R(p)$ and  $S_p[b^*,a]$, using the operator relation \eqref{out=in_q}.
As to $S_p[b^*, a]$ it is convenient to indicate the modes numbers and write it as $S_p(b^*_k, a_k)$.

From the canonical commutation relations \eqref{canonical CR} follows 
\be\label{e^q}
_{\text{out}}\langle \tilde p, b^*|\text{e}^{-\hat{\tilde{q}}}=\,_{\text{out}}\langle \tilde p-i\hbar, b^* |\,, \qquad
e^{\hat q} |p, a\rangle_{\text{in}}=|p-i\hbar, a\rangle_{\text{in}}~,
\ee
and also the standard properties of the coherent states \eqref{CS}
\be\ba\label{Coherent states=}
 &\hat a_{k}\,|p, a\rangle =a_k\,|p, a\rangle\,, \quad &\hat a_k^\dag\,|p, a\rangle =\f{\hbar k}{2}\,\f{\p}{\p a_k}\,|p, a\rangle\,, \\[1mm]
&\langle  b^*, \tilde p|\,\hat b_k^\dag=b_m^*\,\langle  b^*, \tilde p|\,, \quad &\langle  b^*, \tilde p|\,\hat b_{k}=\f{\hbar k}{2}\,\f{\p}{\p b_k^*}\,\,\langle  b^*, \tilde p|\,.
\ea\ee
Projecting the operator relation \eqref{out=in_q} between the coherent states \eqref{CS}, we find for the left hand side
\be\label{L matrix element}
_{\text{out}}\langle \tilde p,\,b^*|:e^{-\hat\Phi_{\text{out}}(x,\bar x)}:|p, a\rangle_{\text{in}}=\d(\tilde{p}+p-i\hbar)\,e^{(p-i\hbar/2)\tau}\,R(p)\,e^{-b^*(x)}
S_p\left(b_k^*-\tfrac{i\hbar}{2}\,e^{-ikx}, a_k\right).
\ee
To calculate  the right hand side one has to use \eqref{e^q}-\eqref{Coherent states=} together with the exchange relation
\be\label{exch-relation}
e^{-\hat a(x)}\,e^{2\hat a^\dag(x+y)}=e^{2\hat a^\dag(x+y)}\,e^{-\hat a(x)}
\left(1-e^{iy}\right)^\hbar.
\ee
This leads to
\be\ba\label{R matrix element}
\d(\tilde{p}+p-i\hbar)\,e^{(p-i\hbar/2)\tau}\,\f{\m^2_q\,R(p-i\hbar)}{4\sinh(\pi p)\,\sinh(\pi p-i\pi\hbar)}\,
e^{-a(x)}\int_0^{2\pi}d\bar{y}\,e^{(p-i\hbar)(\bar{y}-\pi)}\left(1-e^{i\bar y}\right)^\hbar\times\\[1mm]
\int_{0}^{2\pi}dy\,e^{(p-i\hbar)(y-\pi)}\left(1-e^{iy}\right)^\hbar
e^{2a(x+y)}
S_{p-i\hbar}\left(b_k^*,a_k+\tfrac{i\hbar}{2}\,e^{ikx}-i\hbar\,e^{ik(x+y)}\right).
\ea\ee

For vanishing non-zero modes, $a_k=b^*_k=0$, we obtain 
from \eqref{L matrix element} and \eqref{R matrix element} the following equation for $R(p)$:
\be\label{eq R(p)}
R(p)=\f{\pi^2\m^2_q\,\g_p^2}{\sinh(\pi p)\,\sinh(\pi( p-i\hbar))}\,R(p-i\hbar)~,
\ee
where $\g_p$ is the integral
\be\label{Integral}
\g_p:=\int_{0}^{2\pi}\f{dy}{2\pi}\,e^{(p-i\hbar)(y-\pi)}\left(1-e^{iy}\right)^\hbar=\f{\Gamma(1+\hbar)}{\Gamma(1-ip)\Gamma(1+\hbar+ip)}~.
\ee
Using in addition the identity $\sinh(\pi p)\,\Gamma(1+ip)\,\Gamma(1-ip)=\pi p,$
\eqref{eq R(p)} can be written as
\be\label{eq R(p)_1} 
R(p)=\m^2_q\,\f{\Gamma(1+ip)\Gamma(1-\hbar-ip)\Gamma^2(1+\hbar)}{\Gamma(1-ip)\Gamma(1+\hbar+ip)p(p-i\hbar)}\,R(p-i\hbar)~.
\ee
Its solution is 
\be\label{R(p)=}
R(p)=-\left(\m_q^2\,\Gamma^2(\hbar)\right)^{-\f{ip}{\hbar}}\,\f{\Gamma(ip/\hbar)\Gamma(ip)}{\Gamma(-ip/\hbar)\Gamma(-ip)}~.
\ee
This is the unique solution if we require that it has the correct semiclassical limit 
$R=\exp(i/\hbar F^{(0)})$ with $F^{(0)}=2p[\log(p/\mu)-1]$.  
The reflection amplitude corresponds to the 2-point function of Liouville theory \cite{Dorn:1994xn, Zamolodchikov:1995aa}. 

The equation for $S_p(b_k^*,a_k)$ then reduces to
\be\ba\label{eq S_p} 
e^{-b^*(x)}\,S_p\left(b_k^*-\tfrac{i\hbar}{2}\,e^{-ikx}, a_k\right)=
\f{e^{-a(x)}}{\g_p}\int_{0}^{2\pi}\f{dy}{2\pi}\,e^{(p-i\hbar)(y-\pi)}\left(1-e^{iy}\right)^\hbar
e^{2a(x+y)}\times\\[1mm]
S_{p-i\hbar}\left(b_k^*,a_k+\tfrac{i\hbar}{2}\,e^{ikx}-i\hbar\,e^{ik(x+y)}\right).
\ea\ee
We expand $S_p(b^*_k,a_k)$ in powers of the modes as
\be\ba\label{S_p=}
S_p(b^*,a)=1+S_{-1,1}(p)\,b_1^* a_1+S_{-2,2}(p)\,b_2^*a_2+S_{-1,-1,2}(p)\,b_1^{*\,2}a_2+~~~~~~~~\\[1mm] 
S_{-2,1,1}(p)\, b_2^* a_1^2+S_{-1,-1,1,1}(p)\, b_1^{*\,2} a_1^2 +\cdots ~,
\ea\ee
where the $p$-dependent coefficients define the transition amplitudes between the Fock space states, 
which are specified by their indices and one can read them off from  \eqref{CS}. As a result of energy conservation 
the transition coefficients are non-zero only when the sum of the indices is zero \cite{JT2}. 

The first and the second level transition amplitudes were obtained in \cite{Zamolodchikov:1995aa}, using the conformal symmetry of the $S$-matrix.
The same results were reproduced by a slightly different method in \cite{JT2}, where some higher level transitions were also computed.
Here we demonstrate how the transition amplitudes can be obtained from \eqref{eq S_p}.
Setting in \eqref{eq S_p} $b_k^*=0$, for all $k>0$, and comparing the terms linear in $a_1$, we find the relation
\be\label{eq a_1}
\f{i\hbar}{2}\, S_{-1,1}(p)=i+2i\f{\g_{p-i}}{\g_p}\,,
\ee
where $S_{-1,1}(p)$ is the first level transition amplitude in \eqref{S_p=}. Using \eqref{Integral} this yields
\be\label{f_11}
S_{-1,1}(p)=\f{2}{\hbar}\,\f{1+\hbar-ip}{1+\hbar+ip}~,
\ee
which coincides with the result obtained in \cite{Zamolodchikov:1995aa, JT2}. 
The same expression is obtained by setting $a_k=0$ in \eqref{eq S_p} and comparing the terms linear in $b_1^*$.

One can repeat the same procedure at higher levels. At the second level one 
can still obtain all amplitudes by studying eq.\eqref{eq S_p}  at $a=0$ or at $b^*=0$, if one 
combines it with time-reversal symmetry which relates e.g. $S_{-2,1,1}$ to $S_{-1,-1,2}$. 
Doing this one finds the known results. At higher levels one also needs to consider \eqref{eq S_p} 
for $a$ and $b^*$ both nonzero. In this case one finds more complicated equations 
which relate matrix elements with shifted momenta. We will not study these equations here 
as our main goal is to solve eq.\eqref{eq S_p}.

\section{Normal symbol of the $S$-matrix}

In this section we introduce the normal symbol\footnote{Given the 
operator $:\!\hat U(\hat a^\dagger, \hat a,)\!\!:$, whose matrix elements between coherent 
$in$-states reproduce those  of $\hat S$,  we have
$_{\text{in}}\!\langle b^*|:\!\hat U(\hat a^\dagger,\hat a)\!:|a\rangle_{\text{in}}=\,
_{\text{in}}\!\langle b^*|a\rangle_{\text{in}}\,U(b^*,a)=S(b^*,a)$ with 
$_{\text{in}}\!\langle b^*|a\rangle_{\text{in}}=\exp(\frac{2}{\hbar}\sum_{k>0}\frac{b_k^*\,a_k}{k})$.
$U(a^*,a)$ is called the normal symbol of $\hat S$ and  
the operator $:\!\hat U\!:$ can be reconstructed from it.} for the $S$-matrix operator defined in \eqref{S-matrix}. 
This operator describes Fock space transition amplitudes and it contains the momentum $p$ as a parameter. 
For discrete momenta $p=i\hbar N$, where $N$ is a positive integer, we will construct the normal symbol of this operator 
as a multiple  contour integral. In this way we
recover the representation which we had previously derived in \cite{JT3} from a 
conjectured functional integral representation of the generating functional of $S$-matrix elements, 
thus proving its validity, at least for these infinitely many discrete momentum values.  

To this end we introduce the functional
\be\label{U_p}
U_p[b^*,a]=\exp\left(-\f{2}{\hbar}\sum_{k>0}\f{b_k^*\,a_k}{k}\right) S_p[b^*,a]~.
\ee
The function $U_p(b_k^*,a_k)$ is also holomorphic in the modes 
$(b_k^*,\,a_k)$. At $b^*_k=a_k^*$ it coincides with the normal symbol of the $S$-matrix which 
we denote by  $U_p(a_n)$, with $n\neq 0$.

From \eqref{eq S_p} we obtain the following equation for the function $U_p(b^*_k,a_k)$ 
\be\ba\label{eq U_p}
U_p(b^*_k,a_k)=\f{1}{\g_p}\int_0^{2\pi}\f{dy}{2\pi} e^{(p-i\hbar)(y-\pi)}
e^{2[a(x+y)+b^*(x+y)]}
\times~~~~~~~~~~~~~~~~~~~
\\[1mm]
U_{p-i\hbar}\left(b_k^*+\tfrac{i\hbar}{2}\,e^{-ikx},a_k+\tfrac{i\hbar}{2}\,e^{ikx}-i\hbar\,e^{ik(x+y)}\right).
\ea\ee
where we have shifted the $b_k^*$ modes by $b^*_k\mapsto b^*_k+\f{i\hbar}{2}\,e^{-ikx}$.

At $p=i\hbar N $, where $N$ is a positive integer, from \eqref{eq U_p} follows 
\be\ba\label{eq U_N}
U_{i\hbar N}(a_n)=\f{1}{\g_{i\hbar N}}\int_0^{2\pi}\f{dy}{2\pi} e^{i(N-1)\hbar(y-\pi)}
\exp\left(2i\sum_{n\neq 0}\f{a_n}{n}e^{-in(x+y)}\right)
\times
\\[1mm]
U_{i\hbar(N-1)}\left(a_n+\tfrac{i\hbar}{2}\,e^{inx}-i\hbar\,\th_n\,e^{in(x+y)}\right),
\ea\ee
where $\th_n=0$, for $n<0$ and $\th_n=1$, for $n>0$. 
This can be treated as a recursive relation which we will solve. 

For this we assume that $U_0(a_n)=1$, i.e. at $p=0$ there is no scattering. 
This assumption is physically reasonable; it is supported by the classical picture  (see the remark at the end of Section 2.1)
and by the known exact quantum results  at low levels.  
From \eqref{eq U_N} and \eqref{Integral}, we then find
\be\label{U_1=}
U_{i\hbar}(a_n)=\int_0^{2\pi}\f{dy}{2\pi}\,\exp\Big(2i\sum_{n\neq 0}\f{a_n}{n}e^{-in(x+y)}\Big)
=1+\sum_{\n\geq 2}\f{(2i)^\n}{\n!}\sum_{n_1,\dots n_\n}\f{a_{n_1}\dots a_{n_\n}}{n_1\cdots n_\n}\,\d_{n_1+\dots +n_\n}.
\ee 
As expected, the right hand side is $x$-independent and agrees
with the result which  was obtained in  \cite{JT3} from the path integral representation of the $S$-matrix. 

To proceed with the analysis of the recursive relations it is convenient to introduce complex 
variables on the unit circle: $\xi=e^{iy}$, $\z=e^{ix}$ and rewrite \eqref{eq U_N} as
\be\ba\label{U_N=}
U_{i\hbar N}(a_n)
=\f{e^{-i\pi\hbar(N-1)}}{\g_{i\hbar N}}
\oint\f{d\xi}{2\pi i\xi} \, \xi^{\hbar(N-1)}
\exp\left(2i\sum_{n\neq 0}\f{a_n}{n}(\xi\z)^{-n}\right)
\times
\\[1mm]
U_{i\hbar(N-1)}\left(a_n+\tfrac{i\hbar}{2}\,\z^{n}-i\hbar \,\th_n(\xi\z)^{n}\right),
\ea\ee
Then, at $N=2$, from \eqref{U_1=} and \eqref{U_N=} we find 
\be\ba\label{U_2=}
U_{2i\hbar}(a_n)=\f{e^{-i\pi\hbar}}{\g_{2i\hbar}}
\oint\f{d\xi}{2\pi i\xi} \xi^{\hbar} 
\exp\left(2i\sum_{n\neq 0}\f{a_n}{n}(\xi\z)^{-n}\right)
\times
\\[1mm]
\oint\f{d\z_1}{2\pi i\z_1} \exp\left(2i\sum_{n\neq 0}\f{a_n}{n}\z_1^{-n}\right) 
\left(1-\f{\z}{\z_1}\right)^{\hbar}\left(1-\f{\z_1}{\z}\right)^{-\hbar}\left(1-\f{\z\xi}{\z_1}\right)^{-2\hbar},
\ea\ee
with $\z_1=\xi_1\z$. Introducing a new contour variable
$\z_2=\xi\z$, from \eqref{U_2=} one obtains
\be\ba\label{U_2=1}
U_{2i\hbar}(a_n)=\f{e^{-i\pi\hbar}}{\g_{2i\hbar}}
\oint\f{d\z_1}{2\pi i\z_1}\oint \f{d\z_2}{2\pi i\z_2} 
\exp\left(2i\sum_{n\neq 0}\f{a_n}{n}(\z_1^{-n}+\z_2^{-n})\right)\times\\[1mm]
(\z_1\z_2)^{\hbar}\left(\z_1-\z_2\right)^{-2\hbar}.
\ea\ee

The recursive relation \eqref{U_N=} and the calculation similar to \eqref{U_2=} for general $N$ leads to 
\be\ba\label{U_N=2}
U_{i\hbar N}(a_n)=\prod_{\a=1}^{N}\f{1}{\g_{i\hbar\a}}
\oint\f{d\z_1}{2\pi i\z_1}\cdots \oint\f{d\z_N}{2\pi i\z_N}
\exp\left(2i\sum_{n\neq 0}\f{a_n}{n}\,(\z_1^{-n}+\cdots\z_N^{-n})\right)\times\\[1mm]\prod_{1\leq \a<\b\leq N}
(\z_\a\z_\b)^{\hbar}\left(\z_\a-\z_\b\right)^{-2\hbar}.
\ea\ee
This agrees with the expression which we derived in \cite{JT3} from a functional integral representation 
of $S_p(b^*,a)$ which is based on a one-dimensional field theory on a circle. 
It was proposed for arbitrary values of the momenta and was shown to reduce to 
\eqref{U_N=2} for $p=i N \pi$.  
Expanding \eqref{U_N=2}  in powers of $a_n$ one obtains integral representations of the $S$-matrix elements. 
For low values of $N$ one can perform the integrals as we have demonstrated in \cite{JT3}. Here we will not 
attempt to go beyond what was obtained there, as the main purpose of this note is to prove the 
functional integral representation for $S_p(b^*,a)$ introduced in ref.\cite{JT2,JT3}. This was achieved  
for an infinite set of discrete momenta. 

Given the above expression for the normal symbol of the $S$-matrix, we can expllicitly check the assumption 
$U_0(a)=1$. This can indeed be done using the generalization of the Dotsenko-Fateev integrals \cite{DF}
found in \cite{GS} and  the explicit form of $\gamma_{p}$ given in \eqref{Integral}.  Then, within the functional integral 
representation of refs.\cite{JT2,JT3},  the first factor in \eqref{U_N=2} becomes the  
vacuum-to-vacuum amplitude for the one-dimensional field theory, while the second factor, the multiple integral, 
describes the non-trivial scattering.\footnote{We stress that the vacuum-to-vacuum amplitude of Liouville theory is 
given by the reflection amplitude $R(p)$. The vacuum-to-vacuum amplitude we have in mind here is the 
one corresponding to the functional integral representation of $S_p[b^*,a]$ of refs.\cite{JT2,JT3}.} 

Once the $S$-matrix is known for an infinite set of discrete momenta, one can attempt to analytically continued to
physical, i.e. real positive momenta. For this a further generalization of the Dotsenko-Fateev 
integrals is required which is at present not known. What can be obtained, however, is the analytic continuation of  
the vacuum-to-vacuum amplitude to real momenta which leads to a representation in terms of a single integral.  
This is shown in Appendix B. 

\subsubsection*{Acknowledgements}

\noindent
We thank Diptarka Das and Harald Dorn for useful discussions.
G.J. thanks the Max-Planck Institute for Gravitational Physics for warm hospitality during his visits in Golm, 
where essential part of the work was done. This research has been supported by 
Shota Rustaveli National Science Foundation of Georgia (SRNSFG) by the grant [FR-23-17899].

\setcounter{equation}{0}
\def\theequation{A.\arabic{equation}}

\subsection*{Appendix A: Poisson brackets of the chiral fields}

Here we derive the Poisson brackets algebra of the chiral fields using a new form of the
screening charge given by \eqref{A(x)}. The algebra is then used to check the canonicity of the
map \eqref{out=in} and to prepare the system for quantization.

Let us consider the chiral free field
\be\label{chiral FF}
\phi(x)=q+\tfrac{1}{2}\,p\,x+i\sum_{n\neq 0}\f{a_n}{n}\,e^{-inx}\,.
\ee
Its Fourier modes satisfy the canonical relations \eqref{in-canonical PB} and one gets the Poisson brackets
\be\label{PB psi}
\{e^{-\phi(x)},e^{-\phi(y)}\}=\frac{\pi}{2}\,e^{-\phi(x)}\,e^{-\phi(y)}\,\e(x-y)\,.
\ee

As for the Liouville $in$-field, we assume that $p>0$ and introduce the screening charge 
\be\label{A(x)}
A(x)=\int_{-\infty}^xdz\,\,e^{2\phi(z)} .
\ee
This chiral field is quasi-periodic like the exponential $e^{2\phi(x)}$: 
\be\label{A-monodromy}
e^{2\phi(x+2\pi)}=e^{2\pi p}\,e^{2\phi(x)}\,, \qquad A(x+2\pi)=e^{2\pi p}\,A(x)\,.
\ee

We also introduce the chiral field 
\be\label{chi}
\chi(x)=e^{-\phi(x)}\,A(x)\,,
\ee 
and our aim is to derive the Poisson brackets algebra of these chiral fields.

First we calculate the Poisson brackets $\{e^{-\phi(x)},A(y)\}$, which by eqs. \eqref{PB psi}\ and \eqref{A(x)} reads
\be\label{PB psi-A}
\{e^{-\phi(x)},A(y)\}=\pi\,e^{-\phi(x)}\int_{-\infty}^y dz\, \, e^{2\phi(z)}\,\e(z-x)\,.
\ee
Using the relations $A'(z)=e^{2\phi(z)}$ and 
\be\label{eps'}
\e'(z-x)=2\sum_{k=-\infty}^\infty\d(z-x-2\pi k)\,,
\ee
we rewrite the integral term in \eqref{PB psi-A} as follows
\be\label{int A1}
\int_{-\infty}^y dz\,\, A'(z)\e(z-x)=A(y)\e(y-x)-2\int_{-\infty}^y dz\,\, A(z)\sum_{k=-\infty}^\infty \d(z-x-2\pi k)\,.
\ee
The integration of $\d$-functions here, together with \eqref{A-monodromy}, yields
\be\label{int delta}
\int_{-\infty}^y dz\,\, A(z)\sum_{k=-\infty}^\infty \d(z-x-2\pi k)=A(x)\sum_{k=-\infty}^N
e^{2\pi pk}\,,
\ee
where $N$ is defined by the inequalities $y-2\pi<x+2\pi N<y$, i.e. 
\be\label{N=}
2\pi N<y-x<2\pi (N+1)\,. 
\ee
Then $2N+1=\e(y-x)$ and the sum of the geometric progression in \eqref{int delta} is represented as
\be\label{sum=}
\sum_{k=-\infty}^N e^{2\pi pk}=\f{e^{2\pi pN}}{1-e^{-2\pi p}}=\frac{\,e^{\pi
p\,\e(y-x)}}{2\sinh\pi p}\equiv \theta_p(y-x)\,.
\ee
As a result, the Poisson bracket \eqref{PB psi-A} becomes
\be\label{PB psi-A=}
\{e^{-\phi(x)},A(y)\}=\pi\,e^{-\phi(x)}\left[A(y)\,\e(y-x)-2A(x)\,\theta_{p}(y-x)\right].
\ee
Note that \eqref{PB psi-A=} is regular at $y=x$ and one has
\be\label{PB psi(x)-A(x)}
\{e^{-\phi(x)},A(x)\}=-\pi\,\coth(\pi p)\,e^{-\phi(x)}\,A(x)\,.
\ee

Now we calculate the Poisson brackets  $\{A(x),A(y)\}$ given by the double integral
\be\label{PB A-A}
\{A(x),A(y)\}=2\pi\int_{-\infty}^x du\, \, e^{2\phi(u)}\int_{-\infty}^y dz\, \, e^{2\phi(z)}\,\e(u-z)\,.
\ee
Here, the $z$-integration is similar to the one given in \eqref{PB psi-A} and we end up with the $u$-integral
\be\label{PB A-A=1}
\{A(x),A(y)\}=2\pi\int_{-\infty}^x du\, \, e^{2\phi(u)}\left[A(y)\,\e(u-y)+2A(u)\,\th_p(y-u)\right].
\ee
The integration of the first term is again similar to \eqref{PB psi-A} and, together with a partial integration of the second term, we obtain
\be\ba\label{PB A-A=2}
\{A(x),A(y)\}=2\pi \left[ A(x)A(y)\,\e(x-y)-2A^2(y)\,\th_p(x-y)+A^2(x)\,\th_p(y-x)\right]\\
+2\pi\int_{-\infty}^x du\,\,A^2(u)\,\th_p'(y-u).
\ea\ee
Since $\th_p(z)$ is a stair-step function, one gets
\be\label{th'}
\th_p'(y-u)=\sum_{k=-\infty}^\infty e^{-2\pi p k}\,\d(y-u+2\pi k)\,,
\ee
and the integration of the $\d$-functions in the last term of \eqref{PB A-A=2} yields
\be\label{int=}
\int_{-\infty}^x du\,\,A^2(u)\sum_{k=-\infty}^\infty e^{-2\pi p k}\,\d(y-u+2\pi k)=
A^2(y)\sum_{k=-\infty}^M e^{2\pi p k}=A^2(y)\,\th_p(x-y)\,,
\ee
similarly to \eqref{sum=}. This leads to
\be\label{PB A-A=}
\{A(x),A(y)\}=2\pi\left[A(x)A(y)\,\e(x-y)-A^2(y)\, \theta_{p}(x-y)+A^2(x)\,\theta_{p}(y-x)\right].
\ee

Finally, using \eqref{PB psi-A} and \eqref{PB A-A=}, it is straightforward to check that 
\be\label{PB chi}
\{\chi(x),\chi(y)\}=\frac{\pi}{2}\,\chi(x)\,\chi(y)\,\e(x-y)\,.
\ee

We now introduce the anti-chiral free field
\be\label{anichiral FF}
\bar\phi(\bar x)=\bar q+\tfrac{1}{2}\,\bar p\bar x+i\sum_{n\neq 0}\f{\bar a_n}{n}\,e^{-in\bar x}\,,
\ee
where $\{\bar q,\bar p\}=1$ and the non-zero modes satisfy the canonical relations \eqref{in-canonical PB}. The anti-chiral fields $\bar A(\bar x)$ and $\bar\chi(\bar x)$ are
defined as the chiral ones and we similarly get
\be\label{PB anti-chi}
\{\bar\chi(\bar x),\bar\chi(\bar y)\}=\frac{\pi}{2}\,\bar\chi(x)\,\bar\chi(\bar y)\,\e(\bar x-\bar y)\,.
\ee

The Poisson brackets \eqref{PB chi} \eqref{PB anti-chi} are used to compute the Poisson brackets 
for the $out$-field exponentials, which are given as functionals of the $in$-field, cf. \eqref{out=in}. 
One remaining issue is the treatment of the zero modes. 
Note that the $in$-field of Liouville theory \eqref{in-out fields} is given as the sum of the 
chiral \eqref{chiral FF} and the anti-chiral \eqref{anichiral FF} free-fields
\be\label{Phi_in=}
\Phi_{\text{in}}(x,\bar x)=\phi(x)+\bar\phi(\bar x)\,,
\ee
restricted by the constraints $\bar p-p=0$, $\bar q=0$. Therefore, the Poisson bracket for the $in$-field functionals can 
be calculated by the Dirac brackets defined by these two second class constraints.
However, since the coordinate zero mode of $\Phi_{\text{in}}$ is $q+\bar q$ and it has vanishing Poisson brackets with the constraint $\bar p-p$, 
the Dirac brackets of the functions of $\Phi_{\text{in}}$ coincide with their Poisson brackets treated with independent chiral and anti-chiral zero-modes.
Therefore,  the Poisson brackets of the $out$-field exponentials \eqref{PB E's}  directly follow from \eqref{PB chi} and \eqref{PB anti-chi}.

Coming back to the Poisson brackets at coincident points \eqref{PB psi(x)-A(x)}, we obtain 
\be\label{PB 0}
\{e^{-\phi(x)},\sinh(\pi p)\,A(x)\}=0\,.
\ee
We use its quantum analogue in Section 2.1 to fix the operator ordering ambiguities for the construction of the $out$-field exponential \eqref{out=in_q}.

Indeed, due to the monodromy \eqref{A-monodromy}, the chiral field \eqref{A(x)} is represented as
\be\label{A=sinh}
A(x)=\sum_{n=0}^\infty \int_{-2\pi(n+1)}^{-2\pi n}dy\,e^{2\phi(y+x)}=\f{1}{2\sinh(\pi p)}\int_0^{2\pi}dy\,e^{2\phi(y+x)-\pi p}\,,
\ee
and by \eqref{PB 0} one gets
\be\label{PB 0,a}
\{e^{-\phi(x)},\int_0^{2\pi}dy\,e^{2\phi(y+x)-\pi p}\}=0\,.
\ee
Note that for $y\in (0,2\pi)$ one has the vanishing Poisson brackets also with the integrand
\be\label{PB 0,b}
\{e^{-\phi(x)},e^{2\phi(y+x)-\pi p}\}=0\,.
\ee
Its quantum version is used in eq.\eqref{comm=0}, which provides Hermiticity of the $out$-field vertex 
operator in eq.\eqref{out=in_q}. 

\setcounter{equation}{0}
\def\theequation{B.\arabic{equation}}

\subsection*{Appendix B: The vacuum-to-vacuum amplitude}

For a positive integer $N$ we introduce the function
\be\label{Z_N}
Z_N(\hbar)\equiv\prod_{k=1}^N \gamma_{i\hbar k}=\prod_{k=1}^N\f{\G(1+\hbar)}{\G(1+k\hbar)\G(1+\hbar-k\hbar)}\,.
\ee
Our aim is to find its analytical continuation in $N$.

We use the integral representation 
\be\label{log Gamma}
\log\G(z)=\int_0^\infty \f{dt}{t}\left[\f{e^{-zt}-e^{-t}}{1-e^{-t}}+(z-1)e^{-t}\right]\,,
\ee
 and rewrite \eqref{Z_N} as $Z_N(\hbar)=e^{I_N(\hbar)}$, with
 \be\label{I_N(h}
 I_N(\hbar)=\sum_{k=1}^N\int_0^\infty \f{dt}{t}\left[\f{e^{-(1+\hbar)t}+e^{-t}-e^{-(1+k\hbar)t}-e^{-(1+\hbar-k\hbar)t}}{1-e^{-t}}\right]\,.
 \ee
After the summation we get an analytical dependence on $N$ 
\be\label{I_N(h)=}
I_N(\hbar)=\int_0^\infty \f{dt}{t}\left[N\f{e^{-(1+\hbar)t}+e^{-t}}{1-e^{-t}}-\f{{e^{-t}}}{1-e^{-t}}\left(\f{e^{-\hbar t}\left(1-e^{-N\hbar t}\right)}{1-e^{-\hbar t}}+\f{1-e^{N\hbar t}}{1-e^{\hbar t}}\right)\right],
\ee
and for $N\hbar =-ip$ we find
\be\label{I(p,h)=}
I(p,\hbar)=-\f{i}{\hbar}\int_0^\infty \f{dt}{t\left(e^t-1\right)}\,\left[{p}\left(1+e^{-\hbar t}\right)+i\hbar\f{e^{ipt}-e^{-ipt}}{e^{\hbar t}-1}\right].
\ee
The analytical continuation described here is similar to the one used in \cite{Dorn:1994xn} for the 3-point function of Liouville theory.

\end{document}